\documentclass[11pt,a4paper]{article}
\usepackage[T1]{fontenc}
\usepackage{lmodern,amsmath,amssymb,amsthm,microtype}
\usepackage[margin=25mm]{geometry}
\usepackage{xcolor,tikz}
\usetikzlibrary{arrows.meta,decorations.markings}
\usepackage[colorlinks=true,linkcolor=black,citecolor=blue!50!black,urlcolor=blue!50!black]{hyperref}
\hypersetup{pdftitle={An Obstruction to Schwarzschild as a Gauged WZW Target},pdfauthor={Aviv Orly}}
\theoremstyle{plain}
\newtheorem{theorem}{Theorem}
\newtheorem{lemma}{Lemma}
\newtheorem{proposition}{Proposition}
\theoremstyle{definition}
\newtheorem{definition}{Definition}
\numberwithin{equation}{section}
\newcommand{\Tr}{\operatorname{Tr}}
\newcommand{\K}{\mathcal K}
\newcommand{\dd}{\mathrm d}
\newcommand{\ord}{\operatorname{ord}}
\newcommand{\tp}{\mathsf T}
\title{\bfseries An Obstruction to Schwarzschild\\ as a Gauged WZW Target}
\author{Aviv Orly\\[3pt]{\small\itshape School of Physics and Astronomy, Tel-Aviv University, Tel-Aviv 69978, Israel}}
\date{}
\begin{document}
\maketitle
\begin{abstract}
The four-dimensional Schwarzschild black hole cannot be the classical string-frame metric of a regular anomaly-free left--right gauged Wess--Zumino--Witten (GWZW) model of a finite-dimensional group, even locally. The obstruction lies in the analytic behavior of curvature along geodesics. For every such target, the Kretschmann scalar along an affinely parametrized geodesic extends to a single-valued meromorphic function on the whole complex plane of the affine parameter. For Schwarzschild, along a geodesic with nonzero angular momentum and nonconstant radius, it instead has a branch point with exponent $-12/5$. The result holds in Lorentzian and Euclidean signature, including the weakly curved exterior of a large Euclidean black hole, and survives a direct product with a smooth spectator manifold. Nor is Schwarzschild a local $C^2$ limit of such metrics whose parent groups have bounded dimension. This result rules out a broad class of worldsheet constructions for the Schwarzschild black hole and narrows the search for its string-theoretic description.
\end{abstract}

\section{Introduction}

Witten showed that the two-dimensional black hole is an exact gauged Wess--Zumino--Witten (GWZW) background, the coset $SL(2,\mathbb R)/U(1)$ \cite{Witten} (see also \cite{MSW,EFR}), and its exact metric is known \cite{DVV}. Other GWZW backgrounds include the three-dimensional black string $(SL(2,\mathbb R)\times\mathbb R)/\mathbb R$ \cite{HH}, dual to the BTZ black hole \cite{HW,Kaloper}; higher-dimensional cosets $SO(d-1,2)/SO(d-1,1)$ \cite{BS}; cosets with a single time direction, including four-dimensional ones \cite{GQ}; four-dimensional backgrounds from null gaugings \cite{KT}; and a closed expanding universe \cite{NW}. Charged four-dimensional black holes and their near-horizon regions have been described exactly by heterotic cosets and by discrete quotients of WZW models \cite{GPS,Johnson,LS}.

Can four-dimensional Schwarzschild be the target of a gauged WZW model, at least for a large mass $M$ in string units, $\alpha'/M^2\to0$? Here the geometry is held fixed in units of $M$: metrics are compared after rescaling by $M^{-2}$ (Section~\ref{sec:limits}), since otherwise a patch of fixed size in string units at fixed $r/M$ simply becomes flat as $M$ grows. We show that for regular anomaly-free left--right gaugings it cannot, not even as a local $C^2$ limit with bounded group dimension: such gaugings impose a complex-analytic property of curvature that Schwarzschild violates. The Kretschmann scalar of Schwarzschild, continued along a geodesic to complex values of the affine parameter, has a branch point that no regular GWZW target allows.

The proof has three steps. Section~\ref{sec:quotient} shows that the target of a regular GWZW model is locally a metric quotient, by left and right translations, of an auxiliary group with a bi-invariant metric and at most twice the dimension of the WZW group (Proposition~\ref{prop:gwzw}). Section~\ref{sec:lemma1} shows that along any geodesic of such a quotient, polynomial scalar curvature invariants are ratios of exponential polynomials in the affine parameter, and hence extend to single-valued meromorphic functions on the whole complex plane (Lemma~\ref{lem:mero}). Section~\ref{sec:obstruction} finds the branch point of Schwarzschild and concludes (Theorem~\ref{thm:local}). The argument uses only a local metric germ: the contradiction comes from complex continuation, and no physical trajectory has to enter the singularity. Section~\ref{sec:limits} extends the result to limits of bounded dimension, and Section~\ref{sec:outlook} discusses its scope.

\section{Gauged WZW targets}\label{sec:quotient}

Where the gauging is regular, the target of a gauged WZW model is locally a quotient, by left and right translations, of an auxiliary group with a bi-invariant metric (Proposition~\ref{prop:gwzw}). Section~\ref{sec:lemma1} computes curvature along geodesics of such quotients.

\subsection{The target metric}\label{sec:gauged}

We first derive the target metric. Working locally, we use the same letters for Lie algebra and (local) group homomorphisms.

Let the Lie algebra $\mathfrak g$ of a finite-dimensional real Lie group $G$ carry a nondegenerate invariant symmetric form, written $\Tr(XY)$ as is customary, although it need not be a matrix trace. It defines the bi-invariant metric $g_G(v,v')=\Tr(g^{-1}v\,g^{-1}v')$, $v,v'\in T_gG$. By Ado's theorem we may use matrix representatives locally, so that $gXg^{-1}$ is a matrix product for $g\in G$ and $X\in\mathfrak g$.

A left--right gauging is given by a Lie algebra $\mathfrak h$ and two homomorphisms $\ell,\rho:\mathfrak h\to\mathfrak g$, which may embed $\mathfrak h$ differently \cite{BSduality}. Integrated on a Lie group $H$ with Lie algebra $\mathfrak h$, they give the gauge action $g\mapsto\ell(x)\,g\,\rho(x)^{-1}$, $x\in H$. It is anomaly-free when both homomorphisms induce the same form on $\mathfrak h$,
\begin{equation}\label{kappa}
 \kappa(X,Y):=\Tr(\ell X\,\ell Y)=\Tr(\rho X\,\rho Y),\qquad X,Y\in\mathfrak h.
\end{equation}

For $v\in T_gG$, a worldsheet derivative of $g$, the functions
\begin{equation}\label{uw}
 u(X)=\Tr(\ell X\,vg^{-1}),\qquad w(Y)=\Tr(\rho Y\,g^{-1}v)
\end{equation}
pair the right-invariant current $vg^{-1}$ with the left gauge generators and the left-invariant current $g^{-1}v$ with the right ones. The light-cone components $A$ and $B$ of the gauge field are paired by
\begin{equation}\label{Lambda}
 \Lambda_g(X,Y)=\kappa(X,Y)-\Tr(\ell X\,g\,\rho Y\,g^{-1}).
\end{equation}
For the vector gauging of a subgroup $H$ of $G$, $\ell=\rho$ is the inclusion, $\kappa$ is the restriction of $\Tr$, and $\Lambda_g(X,Y)=\Tr(XY)-\Tr(X\,gYg^{-1})$, whose inverse enters the familiar coset metric.

In general, $\Lambda_g$ pairs the effective gauge algebras $\mathfrak h_L=\mathfrak h/\ker\ell$ and $\mathfrak h_R=\mathfrak h/\ker\rho$, and $u\in\mathfrak h_L^*$, $w\in\mathfrak h_R^*$: since $\ker\ell$ and $\ker\rho$ lie in $\ker\kappa$, $\Lambda_g(X,\cdot)$ and $u(X)$ vanish for $X\in\ker\ell$, and $\Lambda_g(\cdot,Y)$ and $w(Y)$ for $Y\in\ker\rho$. Both algebras equal $\mathfrak h$ when $\kappa$ is nondegenerate. We work on the open set where $\Lambda_g$ is invertible, with inverse $\Lambda_g^{-1}:\mathfrak h_L^*\to\mathfrak h_R$.

The gauge fields couple to the currents through $u$ and $w$, and only their classes in $\mathfrak h_L$ and $\mathfrak h_R$ matter. Setting both worldsheet derivatives of $g$ equal to $v$ isolates the symmetric part $\mathcal L_{\rm s}$ of the Lagrangian, which fixes the metric. Up to normalization,
\begin{equation}\label{kinetic}
 \mathcal L_{\rm s}=g_G(v,v)-2u(A)+2w(B)+2\Lambda_g(A,B).
\end{equation}
The equations of motion
\[
 \Lambda_g(\cdot,B)=u,\qquad \Lambda_g(A,\cdot)=-w
\]
give $\Lambda_g(A,B)=u(A)$, so $\mathcal L_{\rm s}=g_G(v,v)+2w(B)$ with $B=\Lambda_g^{-1}u$.

In local coordinates $x^\mu$ on $G$, with bases $X_a$ of $\mathfrak h_L$ and $Y_a$ of $\mathfrak h_R$, let $u_{a\mu}$ and $w_{a\mu}$ be $u(X_a)$ and $w(Y_a)$ for $v=\partial_\mu g$. The resulting degenerate metric on $G$ is \cite{HS}
\begin{equation}\label{metric}
 g_{\mu\nu}=\Tr\bigl(g^{-1}\partial_\mu g\,g^{-1}\partial_\nu g\bigr)+(\Lambda_g^{-1})^{ab}\bigl(w_{a\mu}u_{b\nu}+w_{a\nu}u_{b\mu}\bigr),
\end{equation}
where $(\Lambda_g^{-1})^{ab}$ are the entries of the inverse of the matrix $\Lambda_g(X_a,Y_b)$. For $v=v^\mu\partial_\mu g$ we write $g_{\rm GWZW}(v,v)=g_{\mu\nu}v^\mu v^\nu=g_G(v,v)+2\,w^{\tp}\Lambda_g^{-1}u$.

\begin{definition}[Regular GWZW target]\label{def:gwzw}
A \emph{regular} anomaly-free left--right gauging is a gauging as above, restricted to an open set of $G$ on which $\Lambda_g$ is invertible as a pairing $\mathfrak h_L\times\mathfrak h_R\to\mathbb R$ and the local quotient of $G$ by the kernel of $g_{\rm GWZW}$ is smooth. Its classical target (string frame, leading order in $\alpha'$) is that quotient with the metric induced by \eqref{metric}.
\end{definition}

Where $\Lambda_g$ is invertible, the kernel of $g_{\rm GWZW}$ has constant rank and the local quotient is automatically smooth (end of Section~\ref{sec:quotient2}). Elimination fails where $\Lambda_g$ degenerates, as at the singularity of Witten's Lorentzian black hole \cite{Witten} and on the singular loci of the $SO(2,2)/SO(2,1)$ model of \cite{BSduality}.

\subsection{The gauged metric as an ordinary metric quotient}\label{sec:quotient2}

Section~\ref{sec:lemma1} computes curvature along geodesics by lifting them to geodesics $d_0e^{\tau a}$ of a group. This needs the target to be an ordinary metric quotient of that group by left and right translations. The group is not $G$ but the larger group $D=G\times\bar H$ of \eqref{D}, where $\bar H$ has Lie algebra $\mathfrak h/\ker\kappa$ and the bi-invariant metric $b$ is $\Tr$ on $G$ and minus the form induced by $\kappa$ on $\bar H$. In $D$, the $b$-orthogonal projection of a tangent vector onto the orbit has the on-shell $(A,B)$ of \eqref{kinetic} as its coefficients, and the squared length of the remainder is the target metric \eqref{metric}. This is where the inverse $\Lambda_g^{-1}$ comes from (Proposition~\ref{prop:gwzw}).

\begin{definition}[Regular metric biquotient]\label{def:biquotient}
Let $D$ be a finite-dimensional real Lie group with a nondegenerate bi-invariant metric $b$, and let a Lie group $S$ act on $D$ by left and right translations, $d\mapsto x\,d\,y^{-1}$, where $(x,y)$ is the image of an element of $S$ under a homomorphism $S\to D\times D$. Suppose that on an open set $\mathcal U\subset D$ the orbits have constant dimension and $b$ induces a nondegenerate metric on them. The local orbit space $\mathcal B$ of $\mathcal U$, with the quotient metric that makes the projection $\mathcal U\to\mathcal B$ a semi-Riemannian submersion (its differential maps the $b$-orthogonal complement of each orbit isometrically onto the tangent space of $\mathcal B$; \cite{ONeill66,ONeill}), is a \emph{regular metric biquotient} of $D$.
\end{definition}

\begin{proposition}\label{prop:gwzw}
The classical target of a regular anomaly-free left--right GWZW model with parent group $G$ is locally isometric to a regular metric biquotient of a Lie group $D$ with $\dim D\le2\dim G$.
\end{proposition}

\begin{proof}
\emph{The auxiliary group.} Invariance of $\kappa$ makes $\ker\kappa$ an ideal. Let $\bar H$ be a group whose Lie algebra $\bar{\mathfrak h}=\mathfrak h/\ker\kappa$ carries the nondegenerate invariant form $\bar\kappa$ induced by $\kappa$. Bars denote the projections to $\bar{\mathfrak h}$ and $\bar H$. On the group
\begin{equation}\label{D}
 D=G\times\bar H\quad\text{with metric}\quad b=\Tr\oplus(-\bar\kappa)
\end{equation}
the metric $b$ is bi-invariant and nondegenerate, but in general indefinite. The group $H\times H$ acts on $D$ by left and right translations,
\begin{equation}\label{action}
 (x,y)\cdot(g,h)=\bigl(\ell(x)\,g\,\rho(y)^{-1},\ \bar x\,h\,\bar y^{-1}\bigr).
\end{equation}
Every orbit locally meets the slice $h=1$, and $\{\bar x=\bar y\}$ is the subgroup that preserves the slice. We identify $T_{(g,1)}D$ with $\mathfrak g\oplus\bar{\mathfrak h}$ by left translation, so that $v\in T_gG$ corresponds to $(g^{-1}v,0)$. Then $(X,Y)\in\mathfrak h\oplus\mathfrak h$ generates the orbit vector
\begin{equation}\label{xi}
 \xi(X,Y)=\bigl(g^{-1}\ell X\,g-\rho Y,\ \overline{X-Y}\bigr).
\end{equation}

\emph{Metric comparison.} By \eqref{xi}, $\xi(X,0)=0$ for $X\in\ker\ell$ and $\xi(0,Y)=0$ for $Y\in\ker\rho$, so $\xi$ descends to $\mathfrak h_L\oplus\mathfrak h_R$. In a basis of $\mathfrak h_L\oplus\mathfrak h_R$, the invariance of $\Tr$ and \eqref{kappa} give the Gram matrix $M$ of the orbit vectors and their pairings $c$ with $(g^{-1}v,0)$; the minus sign of $\bar\kappa$ in $b$ cancels the terms $\kappa(X,X')$ and $\kappa(Y,Y')$:
\begin{align}
 b\bigl(\xi(X,Y),\xi(X',Y')\bigr)
 &=\Tr\bigl[(g^{-1}\ell X\,g-\rho Y)(g^{-1}\ell X'g-\rho Y')\bigr]-\kappa(X-Y,X'-Y')\notag\\
 &=\Lambda_g(X,Y')+\Lambda_g(X',Y),\qquad M=\begin{pmatrix}0&\Lambda_g\\ \Lambda_g^{\tp}&0\end{pmatrix},\label{gram}\\
 b\bigl(\xi(X,Y),(g^{-1}v,0)\bigr)&=\Tr(\ell X\,vg^{-1})-\Tr(\rho Y\,g^{-1}v)\notag\\
 &=u(X)-w(Y),\qquad c=(u,-w).\label{pair}
\end{align}
$M$ is invertible if and only if $\Lambda_g$ is. Then $\xi$ is injective on $\mathfrak h_L\oplus\mathfrak h_R$, and the orbit is nondegenerate of dimension $2\dim\mathfrak h_L$. Nondegeneracy of $b$ on the orbit, not positivity, is what makes $T_{(g,1)}D$ the direct sum of the vertical space, tangent to the orbit, and its $b$-orthogonal complement, the horizontal space. The vertical part of $(g^{-1}v,0)$ is $\xi(X,Y)$ with $(X,Y)=M^{-1}c$, that is, $\Lambda_g(\cdot,Y)=u$ and $\Lambda_g(X,\cdot)=-w$: the equations of motion, so $(X,Y)=(A,B)$ on shell. As $b\bigl((g^{-1}v,0),(g^{-1}v,0)\bigr)=g_G(v,v)$, the horizontal part has squared length
\[
 g_G(v,v)-c^{\tp}M^{-1}c=g_G(v,v)+2\,w^{\tp}\Lambda_g^{-1}u,
\]
which is \eqref{metric}.

\emph{The local isometry.} Let the gauging be regular on $U_G\subset G$. The orbits through $(g,1)$, $g\in U_G$, are then nondegenerate of constant dimension $2\dim\mathfrak h_L$, as are all orbits in their saturation, since the action is isometric. The $\bar H$-components $\bar X$ of $\xi(X,0)$ fill $\bar{\mathfrak h}$, so every tangent vector of $D$ at $(g,1)$ is congruent modulo vertical vectors to some $(g^{-1}v,0)$. Hence the saturation is open, and its local orbit space is a regular metric biquotient of $D$, with $S=H\times H$ in Definition~\ref{def:biquotient}. The slice map $g\mapsto[(g,1)]$ has surjective differential $v\mapsto[(g^{-1}v,0)]$. By the metric comparison it pulls the quotient metric back to $g_{\rm GWZW}$, whose kernel is therefore that of the differential: the $v$ for which $(g^{-1}v,0)$ is vertical,
\begin{equation}\label{kernel}
 v=\ell X\,g-g\,\rho Y,\qquad X-Y\in\ker\kappa,
\end{equation}
which generate the action of the subgroup $\{\bar x=\bar y\}$ that preserves the slice. The slice map thus induces a local isometry from the classical target onto the biquotient.

Taking $X=Y$ in \eqref{kernel} gives the original gauge directions. For nondegenerate $\kappa$ they make up the whole kernel, which is then locally the original $H$ gauge action. For degenerate $\kappa$ the kernel may contain further directions, and Definition~\ref{def:gwzw} uses the quotient by the full kernel. In the fully null case $\kappa=0$, the slice-preserving left and right actions are independent, and the target has dimension $\dim G-\dim\mathfrak h_L-\dim\mathfrak h_R=\dim G-2\dim\mathfrak h_L$, as in the null-gauged models of \cite{KT}. In all cases $\dim D=\dim G+\operatorname{rank}\kappa\le2\dim G$, since $\operatorname{rank}\kappa\le\dim\ell(\mathfrak h)$.
\end{proof}

Regularity is local and generic: since $\Lambda_g$ is analytic in $g$, if it is invertible at one point of a connected $G$, it is invertible on an open dense set. There $\xi$ is injective on $\mathfrak h_L\oplus\mathfrak h_R$, so the kernel \eqref{kernel} has constant rank $2\dim\mathfrak h_L-\operatorname{rank}\kappa$. It is tangent to the orbits of the slice-preserving subgroup, so the local quotient is smooth. Only gaugings with $\Lambda_g$ nowhere invertible are excluded; in them some gauge fields survive as Lagrange multipliers.

\section{Curvature along geodesics}\label{sec:lemma1}

A polynomial scalar curvature invariant is a finite linear combination of complete contractions of products of the Riemann tensor with the inverse metric, such as the Kretschmann scalar $\K=R_{abcd}R^{abcd}$. Along geodesics of regular metric biquotients (Definition~\ref{def:biquotient}), such invariants extend meromorphically in the affine parameter $\tau$ to $\mathbb C$ (Lemma~\ref{lem:mero}), unlike $\K$ of Schwarzschild (Section~\ref{sec:obstruction}). We work locally, in any signature.

Exponential polynomials and their orders are
\begin{equation}\label{exppoly}
 F(\tau)=\sum_\lambda p_\lambda(\tau)\,e^{\lambda\tau},\qquad \ord F=\sum_\lambda\bigl(1+\deg p_\lambda\bigr),
\end{equation}
finite sums with distinct $\lambda\in\mathbb C$ and nonzero polynomials $p_\lambda$. They are entire, and the empty sum $F=0$ has order zero.

\begin{lemma}[Meromorphy]\label{lem:mero}
Along any affine geodesic of a regular metric biquotient of $D$, $\K=F/Q$, where $F$ and $Q$ are exponential polynomials, $Q\not\equiv0$, and $\ord F$ and $\ord Q$ are bounded by a constant that depends only on $\dim D$. Hence $\K$ extends to a meromorphic function of $\tau$ on the whole complex plane. Each fixed polynomial scalar curvature invariant has such a representation too, with an order bound that may also depend on the invariant.
\end{lemma}

\begin{proof}
Along the horizontal lift $d_0e^{\tau a}$ of the geodesic to $D$, the vertical fields, the horizontal projector and hence the curvature are rational in a matrix $Z(\tau)$ of exponential polynomials.

\emph{Horizontal lift.} By Definition~\ref{def:biquotient}, $\pi:\mathcal U\to\mathcal B$ is a semi-Riemannian submersion. At $d\in\mathcal U$ the vertical space $\mathcal V_d$, tangent to the orbit, is nondegenerate by regularity, and $\dd\pi$ maps its $b$-orthogonal complement, the horizontal space $\mathcal H_d$, isometrically onto $T_{\pi(d)}\mathcal B$. Geodesics of $\mathcal B$ lift locally to horizontal geodesics $d(\tau)$ of $D$ \cite{ONeill66,ONeill}, which stay horizontal: left and right translations preserve $b$, so the orbit generators $\xi$ are Killing fields with conserved momenta $b(\dot d,\xi)$ that stay zero. As $\nabla_XY=\frac12[X,Y]$ on left-invariant fields, geodesics of $D$, lifts included, are $d(\tau)=d_0e^{\tau a}$ with $d_0\in D$ and $a$ in the Lie algebra $\mathfrak d$ of $D$ (cf.~\cite{BS}).

\emph{The matrix $Z$.} A left-invariant frame $e_i$, in which $b$ is constant, identifies $T_dD$ with $\mathfrak d$. There $d\mapsto e^{tX}d$ and $d\mapsto de^{-tY}$ have generators $d^{-1}Xd$ and $-Y$. With $Z$ the linear map $X\mapsto d^{-1}Xd$ of $\mathfrak d$, the orbit generators are $ZX-Y$, where $(X,Y)\in\mathfrak d\oplus\mathfrak d$ runs over the image of the Lie algebra of $S$. Along the lift $\dot ZX=-[a,ZX]$, so
\begin{equation}\label{Z}
 Z(\tau)X=e^{-\tau a}\bigl(d_0^{-1}Xd_0\bigr)e^{\tau a},\qquad Z(\tau)=e^{-\tau[a,\cdot\,]}Z(0),
\end{equation}
and $Z(\tau)$ carries all the $\tau$-dependence of the orbit generators.

\emph{Horizontal projector.} Near a point of the lift, pairs $(X_k,Y_k)$ for which the $ZX_k-Y_k$ form a basis of the vertical space give the columns of an $n\times m$ matrix $V$, affine in $Z$, with $n=\dim D$ and $m\le n$ the orbit dimension. As only such a basis is used, the action need not be effective. The Gram matrix $V^{\tp}bV$ is invertible by regularity, and
\begin{equation}\label{P}
 P=1-V\,(V^{\tp}bV)^{-1}V^{\tp}b
\end{equation}
projects $b$-orthogonally onto the horizontal space, removing the gauge (vertical) part: $PV=0$, $P\eta=\eta$ if $V^{\tp}b\eta=0$, and $bP$ is symmetric. It is rational in $Z$, and so are its frame derivatives, since the derivative of $ZX=d^{-1}Xd$ along $e_i$ is $-[e_i,ZX]$.

\emph{Curvature.} In this frame the curvature of $D$ is constant, $R^D(X,Y)W=-\frac14[[X,Y],W]$. For horizontal vector fields $U$ and $W$, O'Neill's formula \cite{ONeill66,ONeill} reads
\begin{equation}\label{oneill}
 R^{\mathcal B}(U,W,W,U)=R^D(U,W,W,U)+\tfrac34\,b((1-P)[U,W],(1-P)[U,W]),
\end{equation}
where the left side is evaluated on $\dd\pi\,U$ and $\dd\pi\,W$, $R(U,W,W,U)=b(R(U,W)W,U)$, and $[U,W]$ is the vector-field bracket. The vertical part $(1-P)[U,W]$ plays the role of a Kaluza--Klein field strength, and the last term of \eqref{oneill} is the analogue of the $F^2$ term of Kaluza--Klein reduction. The formula is usually stated for horizontal fields that project to fields on $\mathcal B$, but $(1-P)[U,W]$ at a point depends only on the values of the horizontal fields $U$ and $W$ there, so it also holds for $U=Pe_i$ and $W=Pe_j$, which need not project. Their bracket involves frame derivatives of $P$, so \eqref{oneill} is rational in $Z$. Polarization gives the curvature tensor of $\mathcal B$ on horizontal vectors, and contraction with the horizontal inverse metric $Pb^{-1}$ makes every polynomial scalar curvature invariant rational in $Z(\tau)$.

\emph{Meromorphy.} The entries of $e^{-\tau[a,\cdot\,]}$, hence of $Z(\tau)$, are exponential polynomials with exponents among the eigenvalues of $-[a,\cdot\,]$ and degrees below its Jordan block sizes. Clearing denominators gives $\K=F/Q$ with exponential polynomials $F$ and $Q$, where $Q$ is a power of $\det(V^{\tp}bV)$, nonzero on the short interval where the chosen columns of $V$ form a basis.

The orbit distribution is analytic of constant rank, so the quotient metric and $\K(\tau)$ are real-analytic on the connected real interval $I$ of the geodesic. Thus $Q\K=F$ on all of $I$, and zeros of $Q$ there, where the chosen basis fails, are removable in $F/Q$. As $F$ and $Q$ are entire, $F/Q$ continues $\K$ single-valuedly and meromorphically to $\mathbb C$, without a global complex quotient or lift.

\emph{Order bound, used only in Section~\ref{sec:limits}.} Jordan blocks and matrices have size at most $n$, and orders obey $\ord(F_1+F_2)\le\ord F_1+\ord F_2$ and $\ord(F_1F_2)\le\ord F_1\,\ord F_2$. A fixed invariant needs a fixed number of algebraic operations and frame derivatives, so its rational degrees, hence $\ord F$ and $\ord Q$, are bounded in terms of $n$ and the invariant, for $\K$ in terms of $n$ alone.
\end{proof}

\section{The obstruction}\label{sec:obstruction}

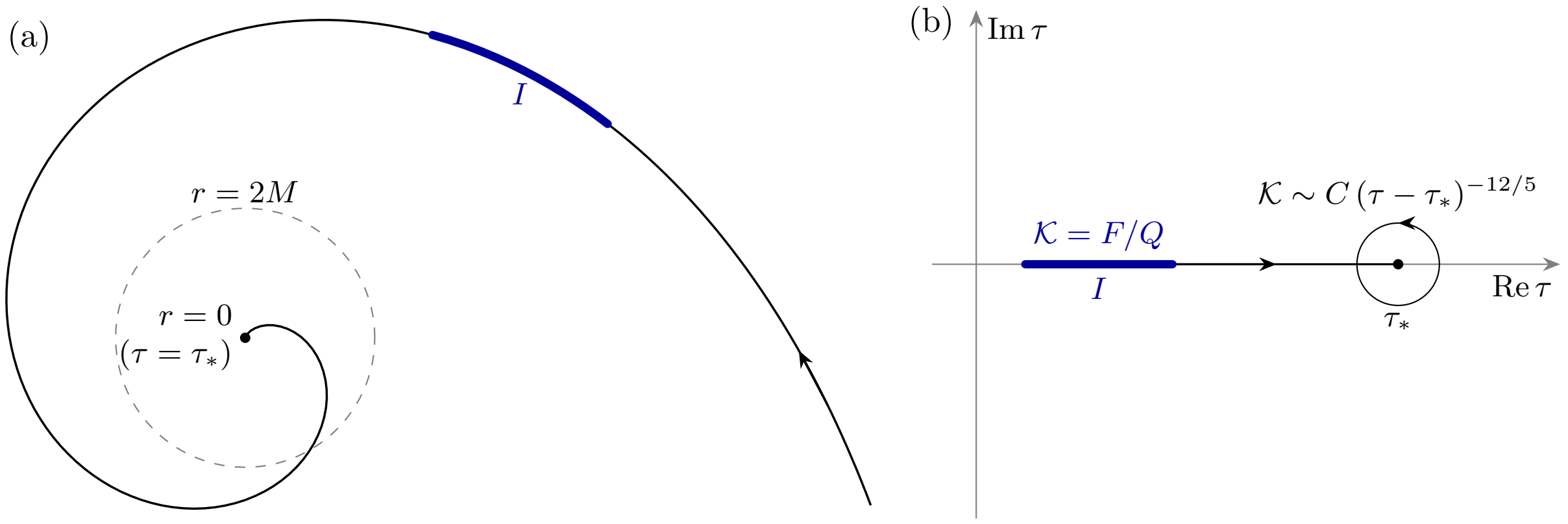
\begin{figure}[t]
\centering
\begin{tikzpicture}[font=\footnotesize, >={Stealth[length=5pt,width=4pt]}]
\colorlet{figaccent}{blue!60!black}
\begin{scope}[x=0.66cm,y=0.66cm]
\draw[gray,dashed,thin] (0,0) circle[radius=2];
\node[anchor=south,inner sep=1.5pt] at (0,2) {$r=2M$};
\draw[line width=0.65pt,line join=round] plot coordinates {
    (9.659,-2.588) (9.589,-2.404) (9.516,-2.220) (9.442,-2.037) (9.365,-1.856) (9.287,-1.675)
    (9.207,-1.495) (9.125,-1.316) (9.041,-1.139) (8.954,-0.962) (8.866,-0.787) (8.776,-0.613)
    (8.683,-0.440) (8.589,-0.268) (8.492,-0.098) (8.393,0.070) (8.292,0.238) (8.188,0.403)
    (8.083,0.567) (7.975,0.729) (7.864,0.890) (7.752,1.049) (7.637,1.205) (7.520,1.360)
    (7.400,1.513) (7.279,1.664) (7.154,1.812) (7.028,1.959) (6.899,2.102) (6.768,2.244)
    (6.634,2.383) (6.498,2.519) (6.360,2.653) (6.219,2.784) (6.076,2.912) (5.930,3.037)
    (5.783,3.159) (5.633,3.277) (5.480,3.393) (5.326,3.505) (5.169,3.613) (5.011,3.718)
    (4.850,3.820) (4.687,3.917) (4.522,4.011) (4.355,4.100) (4.186,4.186) (4.015,4.267)
    (3.843,4.344) (3.669,4.417) (3.494,4.485) (3.317,4.548) (3.138,4.607) (2.958,4.660)
    (2.778,4.709) (2.596,4.753) (2.413,4.791) (2.230,4.825) (2.045,4.853) (1.861,4.875)
    (1.676,4.892) (1.491,4.904) (1.305,4.910) (1.120,4.910) (0.936,4.904) (0.752,4.893)
    (0.568,4.876) (0.385,4.852) (0.204,4.823) (0.024,4.788) (-0.155,4.747) (-0.332,4.700)
    (-0.507,4.647) (-0.680,4.588) (-0.850,4.523) (-1.018,4.452) (-1.183,4.375) (-1.344,4.292)
    (-1.503,4.204) (-1.657,4.109) (-1.808,4.010) (-1.955,3.904) (-2.097,3.793) (-2.235,3.677)
    (-2.368,3.556) (-2.496,3.430) (-2.619,3.299) (-2.736,3.163) (-2.847,3.023) (-2.953,2.878)
    (-3.052,2.730) (-3.144,2.577) (-3.230,2.421) (-3.310,2.262) (-3.382,2.100) (-3.447,1.935)
    (-3.504,1.767) (-3.554,1.597) (-3.596,1.426) (-3.630,1.253) (-3.657,1.079) (-3.675,0.904)
    (-3.685,0.729) (-3.687,0.553) (-3.681,0.378) (-3.666,0.204) (-3.643,0.031) (-3.611,-0.140)
    (-3.571,-0.309) (-3.523,-0.476) (-3.467,-0.640) (-3.402,-0.801) (-3.330,-0.957) (-3.250,-1.110)
    (-3.161,-1.257) (-3.066,-1.400) (-2.963,-1.537) (-2.853,-1.668) (-2.736,-1.792) (-2.613,-1.910)
    (-2.484,-2.020) (-2.349,-2.122) (-2.209,-2.216) (-2.064,-2.302) (-1.915,-2.379) (-1.761,-2.446)
    (-1.605,-2.504) (-1.445,-2.552) (-1.284,-2.590) (-1.121,-2.617) (-0.957,-2.634) (-0.793,-2.640)
    (-0.630,-2.635) (-0.468,-2.619) (-0.308,-2.592) (-0.151,-2.554) (0.002,-2.506) (0.150,-2.446)
    (0.292,-2.377) (0.429,-2.297) (0.557,-2.208) (0.678,-2.109) (0.789,-2.002) (0.891,-1.886)
    (0.981,-1.764) (1.061,-1.635) (1.127,-1.501) (1.181,-1.363) (1.221,-1.222) (1.248,-1.079)
    (1.260,-0.936) (1.257,-0.794) (1.240,-0.656) (1.208,-0.522) (1.163,-0.394) (1.104,-0.275)
    (1.032,-0.165) (0.950,-0.068) (0.858,0.017) (0.758,0.086) (0.653,0.139) (0.544,0.175)
    (0.437,0.194) (0.333,0.195) (0.236,0.179) (0.152,0.148) (0.083,0.107) (0.034,0.061)
    (0.007,0.019) (0.000,0.000)};
\draw[figaccent,line width=2.4pt,line cap=round] plot coordinates {
    (5.597,3.304) (5.431,3.429) (5.259,3.552) (5.081,3.672) (4.896,3.791) (4.705,3.906)
    (4.507,4.019) (4.301,4.128) (4.087,4.234) (3.865,4.334) (3.635,4.430) (3.396,4.520)
    (3.148,4.603) (2.891,4.679)};
\node[figaccent,anchor=62,inner sep=2.5pt] at (4.405,4.074) {$I$};
\draw[->,line width=0.65pt] (8.991,-1.035) -- (8.548,-0.196);
\fill (0,0) circle[radius=1.5pt];
\node[anchor=east,align=right,inner sep=1pt,xshift=-2.5pt] at (0,0) {$r=0$\\$(\tau=\tau_*)$};
\node[font=\small,anchor=north west,inner sep=0pt] at (-3.687,4.910) {(a)};
\end{scope}
\begin{scope}[xshift=7.0cm,yshift=0.75cm]
\draw[gray,thin,->] (0,0) -- (6.4,0) node[black,anchor=north east,inner sep=2pt,yshift=-1pt] {$\operatorname{Re}\tau$};
\draw[gray,thin,->] (0.45,-2.592) -- (0.45,2.592) node[black,anchor=north west,inner sep=2pt,xshift=1pt] {$\operatorname{Im}\tau$};
\node[font=\small,anchor=north west,inner sep=0pt] at (-0.25,2.642) {(b)};
\draw[semithick] (2.45,0) -- (4.75,0);
\draw[semithick,->] (2.45,0) -- (3.510,0);
\draw[figaccent,line width=2.4pt,line cap=round] (0.95,0) -- (2.45,0);
\node[figaccent,anchor=south,inner sep=2pt] at (1.700,0.06) {$\K=F/Q$};
\node[figaccent,anchor=north,inner sep=2pt] at (1.700,-0.06) {$I$};
\draw[thin,postaction={decorate},decoration={markings,mark=at position 0.25 with {\arrow{>}}}] (4.75,0) circle[radius=0.42];
\fill (4.75,0) circle[radius=1.5pt];
\node[anchor=north,inner sep=2pt] at (4.75,-0.42) {$\tau_*$};
\node[anchor=south,inner sep=2pt] at (4.75,0.46) {$\K\sim C\,(\tau-\tau_*)^{-12/5}$};
\end{scope}
\end{tikzpicture}
\caption{(a)~A Lorentzian timelike equatorial geodesic ($E=1$, $L=3.9M$) from $r=10M$ reaches $r=0$ at finite affine parameter~$\tau_*$. (b)~A local isometry on a patch containing~$I$ would give the meromorphic $\K=F/Q$ there, but continuing to $\tau_*$ gives the branch point~\eqref{branch}: a loop around $\tau_*$ does not return $\K$ to itself. In the Euclidean section real geodesics stay in $r\ge2M$, and $\tau_*$ is reached only by complex continuation.}
\label{fig:geodesic}
\end{figure}

We now show that Schwarzschild violates Lemma~\ref{lem:mero}. Near the real axis its Kretschmann scalar is holomorphic in the affine parameter, as for any analytic metric, so local analyticity is not the obstruction. The obstruction is that $\K$ is not single-valued: continued along a geodesic to complex values of the affine parameter, $\K$ has a branch point where the geodesic reaches $r=0$ (Figure~\ref{fig:geodesic}), whereas Lemma~\ref{lem:mero} makes it the restriction of a meromorphic function on the whole complex plane.

\begin{theorem}[Local obstruction]\label{thm:local}
No regular metric biquotient of a finite-dimensional Lie group with a nondegenerate bi-invariant metric, and hence no regular classical anomaly-free left--right GWZW target, is locally isometric to Schwarzschild ($M>0$) in either signature, or to its direct product with any semi-Riemannian spectator manifold.
\end{theorem}

\begin{proof}
Proposition~\ref{prop:gwzw} reduces the claim to regular metric biquotients. Write the Schwarzschild metric as
\begin{equation}\label{schw}
 g_{\rm Schw}=\sigma f\,\dd t^2+f^{-1}\dd r^2+r^2\dd\Omega_2^2,\qquad f=1-\frac{2M}{r},
\end{equation}
with $\sigma=-1$ (Lorentzian) or $\sigma=+1$ (Euclidean); in both cases $\K=48M^2/r^6$. Suppose a regular metric biquotient is isometric to \eqref{schw} on an open patch, which after a rotation of the angular coordinates meets the equatorial plane $\theta=\pi/2$. An equatorial affine geodesic $\gamma$ has conserved $E=f\dot t$ and $L=r^2\dot\phi$, from the Killing vectors $\partial_t$ and $\partial_\phi$, and constant $s=g_{\rm Schw}(\dot\gamma,\dot\gamma)$, the squared norm of its tangent vector, which is negative, zero or positive for timelike, null or spacelike geodesics in Lorentzian signature, so that
\begin{gather}
 \dot r^2=f\Bigl(s-\frac{L^2}{r^2}\Bigr)-\sigma E^2\;\sim\;\frac{2ML^2}{r^3}\qquad(r\to0,\ L\neq0),\label{radial}\\
 p(r):=r^3\dot r^2=(s-\sigma E^2)\,r^3-2Ms\,r^2-L^2r+2ML^2,\qquad p(0)=2ML^2\neq0\quad\text{for }L\neq0.\label{poly}
\end{gather}
Choose such a geodesic $\gamma$ in the patch with $L\neq0$, so that $p(0)\neq0$, and $\dot r\neq0$ on a real interval $I$; any equatorial initial velocity with $\dot r\neq0$ and $\dot\phi\neq0$ gives one (Figure~\ref{fig:geodesic}). Through the isometry it is an affine geodesic of the biquotient, with the same function $\K(\tau)$, so by Lemma~\ref{lem:mero}
\begin{equation}\label{QKF}
 Q\,\K=F\quad\text{on }I,
\end{equation}
for exponential polynomials $F$ and $Q\not\equiv0$.

\emph{Continuation to $r=0$.} Put $\zeta=\sqrt r>0$ on $I$. There, by \eqref{poly}, $\dd\tau/\dd\zeta=2\zeta^4p(\zeta^2)^{-1/2}$ with $p(\zeta^2)^{1/2}=\zeta^3\dot r$. This square root has local holomorphic branches wherever $p(\zeta^2)\neq0$, including at $\zeta=0$, since $p(0)\neq0$. We continue $\tau(\zeta)$ along a path in the complex $\zeta$-plane from a point of $\sqrt{r(I)}$ to $\zeta=0$ that avoids the zeros of $p(\zeta^2)$. It reaches a finite value $\tau_*=\tau(0)$, and
\begin{gather}
 \tau-\tau_*=c_1\,\zeta^5\bigl(1+O(\zeta^2)\bigr),\qquad c_1=\tfrac25\,p(0)^{-1/2}\neq0,\label{zeta}\\
 r\sim c\,(\tau-\tau_*)^{2/5},\qquad\boxed{\displaystyle \K=\frac{48M^2}{\zeta^{12}}\sim C\,(\tau-\tau_*)^{-12/5}}\,,\label{branch}
\end{gather}
with the bracket holomorphic in $\zeta$, $p(0)^{-1/2}$ taken on the continued branch, and nonzero constants $c$ and $C$. So $\K$ has a branch point at $\tau_*$. The choice $L\neq0$ matters: along a timelike radial geodesic with $E=1$ in Lorentzian signature, $r^{3/2}$ is linear in $\tau$, and $\K\propto(\tau-\tau_*)^{-4}$ has only a pole.

\emph{Contradiction.} To avoid inverting $\tau(\zeta)$, we multiply \eqref{QKF} by $\zeta^{12}$. On $I$ this gives
\begin{equation}\label{cross}
 48M^2\,Q\bigl(\tau(\zeta)\bigr)=\zeta^{12}\,F\bigl(\tau(\zeta)\bigr).
\end{equation}
Both sides are holomorphic along the path, since $F$ and $Q$ are entire, so by the identity theorem \eqref{cross} holds along the whole path, up to $\zeta=0$. Its left side is not identically zero, because $Q\not\equiv0$ and $\tau(\zeta)$ is not constant, so neither is its right side. Let $n_Q$ and $n_F$ be the multiplicities of $\tau_*$ as a zero of $Q$ and of $F$, zero if they do not vanish there. By \eqref{zeta}, the two sides of \eqref{cross} then vanish at $\zeta=0$ with multiplicities $5n_Q$ and $12+5n_F$. Thus $5n_Q=12+5n_F$, which is impossible for integers. No real geodesic needs to reach $r=0$.

\emph{Spectator.} Let $g=g_{\rm Schw}\oplus g_N$, with the spectator metric $g_N$ smooth at a point $x_0\in N$, and take $\gamma$ constant in $N$ at $x_0$. Then $\K=\K_{\rm Schw}+\K_N(x_0)$ along $\gamma$, a finite constant shift, and \eqref{QKF} gives $Q\,\K_{\rm Schw}=F-\K_N(x_0)\,Q$, again with an exponential polynomial on the right, so the same contradiction follows.
\end{proof}

The singularity of $\K$ found in the proof is a branch point, so it cannot come from loci where the gauge fields of a GWZW model fail to be eliminated, such as the singularity of Witten's Lorentzian black hole. Along a geodesic such loci give at most poles of $\K=F/Q$. Indeed, in Lemma~\ref{lem:mero} $Q$ can be a power of the Gram determinant of a fixed basis of effective orbit generators \eqref{xi}, which is $\pm(\det\Lambda_g)^2$ on the slice $h=1$ by \eqref{gram}. As the action is isometric, on the real lift $Q$ vanishes only on orbits through points where $\Lambda_g$ degenerates, and there, as at any complex zero of $Q$, $F/Q$ has a pole or a removable singularity.

\paragraph{Euclidean signature.}
The continuation to $r=0$ is a mathematical test of local equality, not a use of the classical approximation in a high-curvature region. Equality on an exterior patch fixes the analytic germ of $\K(\tau)$, whose continuation has the branch point \eqref{branch}, and a function meromorphic on the whole complex plane cannot have that germ. The Euclidean section makes this explicit. It covers only $r\ge2M$, with $t$ of period $8\pi M$ for smoothness at $r=2M$. There $\K\le\K(2M)=3/(4M^4)$, so the curvature is small in string units everywhere once $M^2\gg\alpha'$. Theorem~\ref{thm:local} still applies: it needs a GWZW description only on an open patch, and it reaches the branch point only by complex continuation, never along a real geodesic. So even this smooth, weakly curved geometry, the four-dimensional analogue of the $SL(2,\mathbb R)/U(1)$ cigar, is not locally isometric to any regular GWZW target. For the cigar itself, $g=k(\dd\rho^2+\tanh^2\!\rho\,\dd\psi^2)$, a radial geodesic $\rho=\rho_0+v\tau$ with $v\neq0$ has $\K=16k^{-2}\operatorname{sech}^4(\rho_0+v\tau)$, meromorphic as Lemma~\ref{lem:mero} requires, with poles of order four.

\section{Limits of bounded dimension}\label{sec:limits}

Could Schwarzschild arise as a limit of regular GWZW targets, even though no single target is locally Schwarzschild? Not if their parent-group dimensions stay bounded (Theorem~\ref{thm:limit}). Lemma~\ref{lem:mero} gives more than meromorphy: a bound on the group dimension bounds the orders of the exponential polynomials. Lemma~\ref{lem:closure} shows that, under such a bound, a uniform limit of ratios $F_n/Q_n$ on a real interval is again a ratio $F/Q$ of exponential polynomials on a subinterval, even though the exponents may diverge and the zeros of the denominators may move. Its proof has three moves: cluster the exponents, keep the clusters of maximal amplitude, and separate fast oscillations by integrating against test functions.

\begin{lemma}[Closure of bounded-order quotients]\label{lem:closure}
If continuous $f_n\to f$ uniformly on a compact real interval of positive length and $Q_nf_n=F_n$ with $F_n,Q_n$ exponential polynomials of uniformly bounded order, $Q_n\not\equiv0$, then $Qf=F$ on an open subinterval for exponential polynomials $F,Q$, $Q\not\equiv0$.
\end{lemma}

\begin{proof}
\emph{Clusters.} By the order bound, on a subsequence the exponents $\lambda_{nj}$ of $F_n,Q_n$ have fixed number and polynomial degrees, and each difference $\lambda_{nj}-\lambda_{nk}$ converges or diverges in modulus. Convergence of differences is an equivalence relation, whose classes are the clusters $\beta$. Factoring out a representative exponent $\mu_{n\beta}$ per cluster gives
\begin{equation}\label{clusters}
 (F_n,Q_n)=\sum_\beta c_{n\beta}\,e^{\mu_{n\beta}\tau}\,(a_{n\beta},b_{n\beta}),
\end{equation}
where $c_{n\beta}>0$ and the exponents of $a_{n\beta},b_{n\beta}$ converge.

The order $\ord F$ is also the least order of a constant-coefficient linear ODE annihilating $F$. Hence $a_{n\beta},b_{n\beta}$ solve a common monic such ODE of fixed order $m_\beta$ whose roots are their exponents, each with multiplicity one more than its largest polynomial degree, so its coefficients converge. At a fixed $\tau_0$ we choose $c_{n\beta}$ so that
\[
 \sum_{j=0}^{m_\beta-1}\bigl(|a_{n\beta}^{(j)}(\tau_0)|^2+|b_{n\beta}^{(j)}(\tau_0)|^2\bigr)=1.
\]
By compactness and continuous dependence on coefficients and data, a subsequence converges uniformly on compact sets to a solution $(a_\beta,b_\beta)$ of the limiting ODE with unit data. These limits are exponential polynomials, not both zero, even if roots coalesce.

\emph{Maximal amplitude.} We write $c_{n\beta}e^{\mu_{n\beta}\tau}=e^{h_{n\beta}(\tau)}e^{i\omega_{n\beta}\tau}$ with log-amplitude and frequency
\begin{equation}\label{amplitudes}
 h_{n\beta}(\tau)=\log c_{n\beta}+\tau\operatorname{Re}\mu_{n\beta},\qquad \omega_{n\beta}=\operatorname{Im}\mu_{n\beta}.
\end{equation}
Each difference $h_{n\beta}-h_{n\beta'}$ is affine in $\tau$. If its intercept and slope stay bounded, it converges on a subsequence. Otherwise, on a subsequence the norm of (intercept, slope) tends to infinity and the difference divided by this norm tends to a nonzero affine function, away from whose zero the difference diverges with fixed sign. On an open subinterval $J$ avoiding these finitely many zeros, every difference converges or diverges with fixed sign, locally uniformly.

Clusters are comparable if their log-amplitude difference converges on $J$. The sign of divergence totally orders the comparability classes. We fix $\beta_0$ in a maximal class $T$ and divide $Q_nf_n-F_n=0$ by $e^{h_{n\beta_0}}$. Clusters outside $T$ carry factors tending to zero, with bounded brackets $b_{n\beta}f_n-a_{n\beta}$, so locally uniformly on $J$
\begin{equation}\label{RL}
 \sum_{\beta\in T}e^{i\omega_{n\beta}\tau}\,e^{\alpha_\beta(\tau)}\,\bigl(b_\beta f-a_\beta\bigr)\longrightarrow0,
\end{equation}
with $\alpha_\beta$ the affine limit of $h_{n\beta}-h_{n\beta_0}$.

\emph{Fast oscillations.} For $\beta\neq\beta'$ in $T$, $\operatorname{Re}(\mu_{n\beta}-\mu_{n\beta'})$ is bounded, being the slope of a convergent affine function, while $\mu_{n\beta}-\mu_{n\beta'}$ diverges, so $|\omega_{n\beta}-\omega_{n\beta'}|\to\infty$. We fix $\beta\in T$, multiply \eqref{RL} by $e^{-i\omega_{n\beta}\tau}$ and integrate against a test function $\chi\in C_c(J)$. By the Riemann--Lebesgue lemma the other terms drop out, leaving
\[
 \int\chi\,e^{\alpha_\beta}\,(b_\beta f-a_\beta)\,\dd\tau=0,
\]
so $b_\beta f=a_\beta$ on $J$. If $b_\beta\equiv0$ then $a_\beta\equiv0$ by analyticity, contrary to the normalization. We take $(F,Q)=(a_\beta,b_\beta)$.
\end{proof}

In the following theorem, a local $C^2$ limit of regular metric biquotient metrics is a $C^2$ limit, on a fixed open set, of metrics $g_n$ each isometric to an open subset of a regular metric biquotient of a group $D_n$, its ambient group.

\begin{theorem}[Limit obstruction]\label{thm:limit}
A local $C^2$ limit of regular metric biquotient metrics with uniformly bounded ambient dimensions cannot be Schwarzschild ($M>0$) in either signature, or its direct product with a spectator metric, even after local $C^3$ diffeomorphisms and nonzero constant rescalings. In particular this holds for families of regular classical anomaly-free left--right GWZW targets with bounded parent-group dimension.
\end{theorem}

\begin{proof}
\emph{Reduction.} Rescaling the bi-invariant metric rescales the biquotient metric alike, absorbing the rescalings. Local $C^3$ diffeomorphisms yield $C^2$ metrics and preserve geodesics and Kretschmann scalars. So let metrics $g_n\to g_{\rm Schw}$ or $g_{\rm Schw}\oplus g_N$ in $C^2$ on a patch, each isometric to a regular metric biquotient of a group $D_n$, with $\sup_n\dim D_n<\infty$.

\emph{Geodesics.} We fix the limiting geodesic $\gamma$ as in Theorem~\ref{thm:local}'s proof: equatorial, with $L\neq0$ and $\dot r\neq0$ on a compact interval $I$, and constant in any spectator. Let $\gamma_n$ be the $g_n$-geodesics with the same initial data. As the limiting metric is nondegenerate, the Christoffel symbols converge in $C^1$ on a smaller compact patch. After shrinking $I$, the geodesic equation gives $\gamma_n\to\gamma$ in $C^1$ on $I$.

Curvature uses at most two metric derivatives, so $f_n(\tau):=\K_{g_n}(\gamma_n(\tau))\to\K(\gamma(\tau))$ uniformly on $I$. Through the local isometries, Lemma~\ref{lem:mero} gives $Q_nf_n=F_n$ on $I$ without analyticity of $g_n$. Here $F_n$ and $Q_n\not\equiv0$ are exponential polynomials with orders bounded in terms of $\dim D_n$, hence uniformly.

\emph{Contradiction.} Lemma~\ref{lem:closure} gives exponential polynomials $F$ and $Q\not\equiv0$ with $Q\K=F$ on an open subinterval $J\subset I$, where still $L\neq0$ and $\dot r\neq0$. The continuation argument of Theorem~\ref{thm:local}, started on $J$, again yields $5n_Q=12+5n_F$, also after subtracting a constant spectator contribution. For GWZW families, Proposition~\ref{prop:gwzw} gives $\dim D_n\le2\dim G_n$, bounded when the parent-group dimensions $\dim G_n$ are.
\end{proof}

The dimension bound enters only through the order bound, which Lemma~\ref{lem:closure} needs: a nowhere-differentiable continuous $f$ is a uniform limit of polynomials $F_n$ with $Q_n=1$, but $Qf=F$ with $Q\not\equiv0$ would make $f$ analytic on a subinterval. Even with bounded orders, $Qf=F$ may hold only on a subinterval: $f_n=\tau/(1+e^{-n\tau})$, with numerator and denominator of order $2$, tends uniformly to $\max(\tau,0)$ with error at most $1/(en)$, and this limit is not analytic at $\tau=0$. Whether Theorem~\ref{thm:limit} itself needs the bound is open.

\paragraph{Large mass.}
This answers the large-mass question of the introduction. As $M\to\infty$, a patch of fixed size in string units at fixed $r/M$ becomes flat, so metrics must be compared in units of $M$. Write $g_{\rm Schw}(M)$ for the Schwarzschild metric of mass $M$. The rescaling $\Phi_M$ of time and radius then gives
\[
 \Phi_M(\bar t,\bar r,\theta,\phi)=(M\bar t,M\bar r,\theta,\phi),\qquad M^{-2}\Phi_M^*g_{\rm Schw}(M)=g_{\rm Schw}(1).
\]
Thus, by Theorem~\ref{thm:limit}, if GWZW targets $g_n$ with bounded parent-group dimension are compared with Schwarzschild of masses $M_n\to\infty$, their normalized metrics $M_n^{-2}\Phi_{M_n}^*g_n$ cannot converge in $C^2$ to $g_{\rm Schw}(1)$ on a fixed open patch. It does not assert an obstruction to the flat limit seen in regions of fixed physical size at fixed $r/M$ as $M\to\infty$.

\section{Outlook}\label{sec:outlook}

The obstruction needs neither spherical symmetry of the candidate nor a classification of groups; it uses only the analytic form of curvature along geodesics. It does not exclude general CFTs, constrained gaugings (gauge fields left as Lagrange multipliers), heterotic cosets in which worldsheet fermions cancel the anomaly of the bosonic gauging \cite{Johnson}, limits of unbounded dimension, unsuppressed quantum corrections, or dilaton-dependent frame changes. Linear sigma models, used in \cite{CMW} to study the black hole/string transition, remain an alternative route.

\section*{Acknowledgments}
I thank Yoav Zigdon for helpful discussions. In preparing this paper I used the AI models Claude Opus 5.5 and Claude Sonnet 5.5 (Anthropic) and GPT-6.2 Sol (OpenAI, through Codex) to check calculations, search the literature and edit the text; I take full responsibility for its content.

\end{document}